\documentclass[a4,useAMS,usenatbib,usegraphicx]{mn2e}
\def\aapr{\ref@jnl{A\&A~Rev.}}		

\include{epsf}
\usepackage{float}
\usepackage{color}
\usepackage{latexsym}
\usepackage{amsmath}
\usepackage{amssymb}
\usepackage{multirow}

\usepackage{epsf}
\epsfclipon

\title[Harassment of globular cluster systems]{The impact of galaxy harassment on the globular cluster systems of early-type cluster dwarf galaxies}
\author[R.Smith et al]
{\parbox{\textwidth}{R.Smith$^{1}$\thanks{E-mail:rsmith@astro-udec.cl},
R.S\'anchez-Janssen$^{2}$, 
M. Fellhauer$^{1}$, 
T. H. Puzia$^{3}$, 
J. A. L. Aguerri$^{4{\rm{,}}5}$ and
J. P. Farias$^{1}$}\vspace{0.4cm}\\
\parbox{\textwidth}{$^{1}$Departamento de Astronomia, Universidad de Concepcion, Casilla 160-C, Concepcion, Chile\\
$^{2}$ European Southern Observatory, Alonso de Cordova 3107, Vitacura, Santiago, Chile
$^{3}$ Department of Astronomy and Astrophysics, Pontifica Universidad Catolica, Santiago, Chile 7820436\\
$^{4}$ Instituto de Astrof\'isica de Canarias, E-38200 La Laguna, Tenerife, Spain\\
$^{5}$ Departamento de Astrof\'isica, Universidad de La Laguna, E-38205 La Laguna, Tenerife, Spain}}
\begin{document}

\date{Accepted to MNRAS 12/11/12}

\pagerange{\pageref{firstpage}--\pageref{lastpage}} \pubyear{2011}

\maketitle

\label{firstpage}

\begin{abstract}
The dynamics of globular cluster systems (GCSs) around galaxies are often used to assess the total enclosed mass, and even to constrain the dark matter distribution. The globular cluster system of a galaxy is typically assumed to be in dynamical equilibrium within the potential of the host galaxy. However cluster galaxies are subjected to a rapidly evolving and, at times, violently destructive tidal field. We investigate the impact of the harassment on the dynamics of GCs surrounding early type cluster dwarfs, using numerical simulations. We find that the dynamical behaviour of the GCS is strongly influenced by the fraction of bound dark matter $f_{\rm{DM}}$ remaining in the galaxy. Only when $f_{\rm{DM}}$ falls to $\sim15\%$, do stars and GCs begin to be stripped. Still the observed GC velocity dispersion can be used to measure the true enclosed mass to within a factor of 2, even when $f_{\rm{DM}}$ falls as low as $\sim3\%$. This is possible partly because unbound GCs quickly separate from the galaxy body. However even the distribution of {\it{bound}} GCs may spatially expand by a factor of 2-3. Once $f_{\rm{DM}}$ falls into the $<3\%$ regime, the galaxy is close to complete disruption, and GCS dynamics can no longer be used to reliably estimate the enclosed mass. In this regime, the remaining bound GCS may spatially expand by a factor of 4 to 8. It may be possible to test if a galaxy is in this regime by measuring the dynamics of the stellar disk. We demonstrate that if a stellar disk is rotationally supported, it is likely that a galaxy has sufficient dark matter, that the dynamics of the GCS can be used to reliably estimate the enclosed mass.
\end{abstract}


\section{Introduction}
\subsection{Cluster galaxies in the hierarchical galaxy formation scenario}
In the current framework of structure formation, cold dark matter (CDM) haloes grow hierarchically by the merging of smaller, virialised haloes. After infall, these subhaloes continue orbiting within the main halo, with tidal interactions stripping their masses from the outside-in and, if massive enough, spiraling in as dynamical friction slows down their velocities.~Thus, depending on their concentration and orbital distribution, subhaloes can lose substantial mass and eventually disrupt or merge at the halo centre. This is a process which is specially efficient in the inner, dense central regions of the main halo.~The interplay between continuous accretion and disruption results in haloes containing $\sim$10-20\% of their mass in the form of self-bound subhaloes (\citealp{Ghigna1998}; \citealp{Gill2004}).
Galaxies form in DM haloes as baryons cool and contract towards the centre of the halo potential well (\citealp{White1978}). In this context, as smaller haloes are accreted and become subhaloes of larger hosts, their galaxies can be identified as a satellite population.~Virialised galaxy clusters conform the paradigm of this scenario, with several hundreds of galaxies orbiting within $10^{14}\!-\!10^{15} {M}_{\odot}$ haloes.~Close, high speed tidal encounters between cluster galaxies are predicted to be frequent, and can result in substantial mass loss -- a process known as `harassment' (\citealp{Moore1996}).~Given that galaxies reside deep within their halo potential well, the efficiencies of DM and baryons stripping are not straightforwardly related (\citealp{Penarrubia2008}), and understanding the fate of satellite galaxies has become one of the greatest challenges for models of galaxy formation (\citealp{Wetzel2010}).

The current situation is somewhat paradoxical.~While theory and numerical simulations provide accurate descriptions of the stripping process and mass evolution of DM subhaloes, the evolution of their baryonic component is, in comparison, more difficult to track. For instance, most semianalytic models have, so far, simply assumed that the stellar mass remains intact until satellites eventually merge with their central galaxy (\citealp{DeLuciaBlaizot2007}). The opposite is true from an observational point of view: while the stellar content of cluster galaxies is routinely measured, very little is known about their true DM content and the properties of their haloes.~The reason for this discrepancy is that traditional mass determination methods face significant challenges when it comes to determining the total DM halo mass. Stellar dynamics studies  are very expensive in terms of telescope time, and therefore have been often limited to the inner ($\lesssim$ $r_{e}$) regions of luminous galaxies -- which are already known to be not DM-dominated. Only recently have observations been able to probe regions as far as $\approx 3\,r_{e}$, but the stellar component still provides the major contribution to the system dynamics (\citealp{Proctor2009}; \citealp{Spolaor2010}). The situation is even more complex in the case of low-mass cluster galaxies, where the intrinsically low surface brightness restrict stellar kinematics measurements to the inner effective radius (e.g., \citealp{Chilingarian2009}; \citealp{Toloba2011}).
Gravitational lensing has been suggested to be the most reliable method for determining the extent and total mass of cluster subhaloes, but the induced shear is so weak that the approach breaks down at the expected masses of dwarf-hosting haloes (\citealp{Limousin2009}).

\subsection{Globular cluster systems as dynamical probes}
Globular cluster systems (GCSs) perhaps provide the only means by which the kinematics of cluster galaxies at large galactocentric radii can be probed. Globular clusters are old, relatively luminous systems exhibiting very extended galactocentric distributions. Furthermore, and contrary to planetary nebulae -- the other stellar population traditionally used as a kinematic tracer of the host halo (e.g., \citealp{Coccato2009}) -- GCs are found in abundance in cluster galaxies across a wide stellar mass range. They are rather populous systems in all Brightest Cluster Galaxies, but are found as well in cluster early-type dwarfs with masses as low as ${M}_{\star} \approx 10^{6.5} {M}_{\odot}$ (\citealp{Miller1998}; \citealp{Miller2007}; \citealp{Peng2008}; \citealp{Georgiev2008}; \citealp{Georgiev2009}; \citealp{Georgiev2010}). They are known to be effective kinematic probes for luminous galaxies, where GCSs have been used to obtain rotation curves and radial velocity dispersion profiles extending well beyond $r\gtrsim3\,r_{e}$ for about a dozen galaxies (\citealp{Puzia2000}; \citealp{Cote2001}; \citealp{Cote2003}; \citealp{Richtler2004}; \citealp{Peng2004}; \citealp{Hwang2008}; \citealp{Romanowsky2009}; \citealp{Schuberth2010}; \citealp{Foster2011}). Moreover, the consistency between these results and those derived from integrated stellar kinematics in the overlapping regions strongly argues in favour of GCSs being true tracers of the old, metal-poor stellar haloes in massive galaxies. 

The low-mass end of the stellar mass function is a much lesser explored
territory. Perhaps surprisingly, it is now well known that the family of
luminous Virgo dEs (${M}_{\star} \gtrsim 2\times10^{8} {M}_{\odot}$) tend to contain populous GCSs (\citealp{Miller2007}; \citealp{Peng2008}). The majority of these massive, GC-rich dEs are found within the nucleated subclass, and their high GC
mass specific frequencies represent a challenge: having spent a
considerable amount of time within the harsh cluster potential, it is
intriguing how have these systems managed to retain such a high
abundance of GCs. The first kinematical study of GCs in a Virgo cluster dwarf galaxy was that of the dE VCC\,1087 by \cite{Beasley2006}. Spectroscopy of 12 GCs out to $r\approx 6\,r_{e}$ in this ${M}_{\star} \approx 3\times10^{9} {M}_{\odot}$ dE galaxy surprisingly show evidence of dynamically significant rotation in the GCS, with $(v/\sigma)^{*} \gtrsim 1.6$. The implied dynamical mass is relatively high (${M}_{dyn}/{M}_{\star} \sim8$), a result later confirmed by \cite{Beasley2009} for two other luminous Virgo dEs\,\footnote{Following previous work (e.g., Boselli et al. 2008), with this definition we refer to any low-mass, gas-free and non-star-forming cluster galaxy. Throughout this paper they will all be termed dE for simplicity.}. Interestingly, the bulk of the dEs stellar bodies do not seem to share the GCS rotation, and these authors suggest that these features might be consistent with evolution through harassment from a late-type disc galaxy. On the other hand, \cite{Ruben2012} point out that the GC abundance of these dEs is too high (factors 2-8) compared to the one expected from such progenitors, and argue that high total masses might be required for dEs to maintain these populous GCSs during their orbital evolution within Virgo. 
It thus appears that kinematical studies of GCSs in dwarf galaxies offer a unique opportunity to determine their total mass content at large radii -- a region currently inaccessible to any other technique and where DM is expected to be dominant, in agreement with those findings. They can help us bridge the gap between the very high ${M}_{dyn}/{M}_{\star}$ ratios characteristic of Local Group dSphs (\citealp{Strigari2008}) and the rather low values (${M}_{dyn}/{M}_{\star} \sim 3$) systematically found from stellar kinematics of cluster and group dEs (\citealp{Toloba2011}; \citealp{Forbes2011}).

However, as previously stated, the orbital evolution of subhaloes within clusters can significantly influence the dynamical status of GCSs. For instance, cluster galaxies are subjected to a rapidly evolving and, at times, violently destructive tidal field that can bring the system out of dynamical equilibrium -- an implicit assumption in all dynamical models. Also high-speed tidal encounters in a cluster are expected to result in significant dark matter mass loss in dwarf galaxies (\citealp{Mastropietro2005}; \citealp{Aguerri2009}; \citealp{Smith2010a}), and the GCS must settle within the new mass distribution. Moreover, the tidal evolution within the cluster can result in some GCs being stripped, no longer tracing the original host halo and biasing results if included within observational samples.

All these effects may significantly impact the implied dE dynamical masses and yet, ironically, harassment is expected to result in substantial dark matter mass loss that may well be detectable by observing GCS dynamics. It is therefore fundamental that the impact of fast close encounters on mass estimates from GCSs is well understood. 

Here we provide such an investigation by studying the impact of cluster harassment on the dynamics of GCSs surrounding dE galaxies, using numerical simulations. In this paper, we focus on GCS dynamics as a means of measuring the enclosed mass. In a future publication we shall fully investigate additional topics such as induced rotation, spatial distribution, and the fate of the stripped GCs. We describe our numerical code, the cluster harassment model, and galaxy model setup in Section~2, we present our results for mass-loss, effects on spatial distribution, and dynamics of the GCSs in Section~3, and we summarise and draw conclusions in Section~4.

\section{Setup}
\subsection{The code}
In this study we make use of `{\sc{gf}}' (\citealp{Williams2001}; \citealp{Williams1998}), which is a Treecode-SPH algorithm that operates primarily using the techniques described in \cite{Hernquist1989}. The dE galaxies we consider are gas-free and non-star-forming. We therefore do not include an SPH component or star formation recipe to our models, and the code operates purely as a gravitational code without considering hydrodynamics. `{\sc{gf}}' has been parallelised to operate simultaneously on multiple processors to decrease simulation run-times. The Treecode allows for rapid calculation of gravitational accelerations. In all simulations, the gravitational softening length, $\epsilon$, is fixed for all particles at a value of 100 pc, in common with the harassment simulations of \cite{Mastropietro2005}. Gravitational accelerations are evaluated to quadrupole order, using an opening angle $\theta_c=0.7$. A second order individual particle timestep scheme was utilised to improve efficiency following the methodology of \cite{Hernquist1989}. Each particle was assigned a time-step that is a power of two division of the simulation block timestep, with a minimum timestep of $\sim$0.5 yrs. Assignment of time-steps for collisionless particles is controlled by the criteria of \cite{Katz1991}. For details of code testing, please refer to \cite{Williams1998}. 

\subsection{Harassment model}
We model the tidal field of a Virgo-like cluster using a two component model; the first is a dynamical tidal field associated with individual `harasser' galaxies, and the second is a static tidal field to represent the rest of the mass of the cluster that is not associated with galaxies - referred to as the background cluster potential herein. The potential field of each individual harasser and the background is represented using the analytical form for an NFW density distribution (\citealp{Lokas2001}):

\begin{equation}
\label{potNFW}
\Psi = -g_cGM_{200}\frac{ln(1+(r/r_s))}{r}
\end{equation}
\noindent where $g_c=1/[{ln(1+c)-c/(1+c)}]$, $c$ is the concentration  parameter, $r_{\rm{s}}$ is a characteristic radial scalelength, and $M_{200}$ is the virial mass.

We fully define the background cluster potential by choosing its virial mass $M_{200}$=1.6$\times$10$^{14}$M$_\odot$, concentration $c$=4 and virial radius $r_{200}$=1100~kpc. This choice of mass is in close agreement with the Virgo cluster model of \cite{Vollmer2001}, and the concentration is typical for cluster-mass halos in CDM simulations (e.g. \citealp{Gill2004}). Now we give each harasser galaxy its own individual NFW potential field. To make this field dynamical, a one-off N-body simulation of a cluster-mass halo is conducted, and the time-evolving coordinates of a fraction of the particles are logged. This log now provides the position-evolution of the centre of each of the harassing haloes, where each individual galaxy potential field will be super-imposed. We require a mass and concentration for each harasser's potential field to be defined. For mass, we use the Schechter function (\citealp{Schechter1976}) with parameters provided in \cite{Sandage1985}, and assume a constant mass-to-light ratio:

\begin{equation}
M_{tot} = \frac{\phi_\star}{L_\star} \left(\frac{M}{L}\right)_K \int_{L_{min}}^{\infty} \left( \frac{L}{L_\star} \right)^{\alpha + 1} \exp \left( -\frac{L}{L_\star} \right) dL
\label{massfunct}
\end{equation}

\noindent
where $\alpha=-1.25$, and $(M/L)_K=20$. The fitting parameters provided in \cite{Sandage1985} are based on the Virgo cluster catalogue (\citealp{VCC}), and are therefore complete down to a minimum galaxy luminosity of $\sim 2.5 \times 10^7 L_\odot$ assuming a distance modulus of 31.0 for Virgo (\citealp{Mei2007}). We resolve all harasser galaxies down to one-hundredth the mass of our standard dwarf model ($10^9 M_\odot$). Completing the integral in Equation \ref{massfunct} produces a total mass in galaxies that is 14.5$\%$ of the total cluster mass. This is in reasonable agreement with the $\sim 10 \%$ percentage of a cluster's mass locked up in sub-structure, found in $\Lambda$CDM simulations - see \cite{Gill2004}. This process produces 733 harassing galaxies in total. The vast majority of harasser galaxies are dwarfs, with $\sim 90 \%$ of the total number of harassers having a mass less than the standard dwarf galaxy model. The maximum mass galaxies produced are $\sim 10^{12} M_\odot$. The value of the mass-to-light ratio chosen is consistent, if not an upper limit for galaxies in this mass-range (see \citealp{Gilmore2007}). The concentration $c$ of each halo is chosen to follow the trend of higher concentration for lower mass objects as found in $\Lambda$CDM simulations. This is achieved by a fit to concentration values with mass found in \cite{Navarro1996} producing $c = -3 \log (M_{halo}(M_\odot)) + 52$. This fit produces $c \sim 15$ for a $6 \times 10^{11}$ M$_\odot$ mass halo, and $c \sim 20$ for a $1 \times 10^{10}$ M$_\odot$ mass halo. The total mass of our model cluster is fixed, whereas real clusters tend to increase their mass by accretion. Therefore we study the effects of harassment over a limited duration of 2.5~Gyr. In this way, the inaccuracy introduced by assuming a fixed cluster mass is also limited. Further properties of the harassment model can be found in \cite{Smith2010a}.

\subsection{Galaxy models}
Our disk galaxy models consist of 3 components; an NFW dark matter halo (\citealp*{Navarro1996}), an exponential disk of stars, and a spheroidal distribution of globular cluster particles following a Hernquist profile. 

\subsubsection{The dark matter halo}
The dark matter halo of the disk galaxy model has an NFW density profile. The NFW profile has the form:
\begin{equation}
\rho(r) = \frac{\rho_0}{(\frac{r}{r_{\rm{s}}})(1+\frac{r}{r_{\rm{s}}})^2}
\label{NFWdensprof}
\end{equation}
\noindent where $\rho_0$ is the central density. Given that the NFW model has a divergent total mass, the profile is typically truncated outside the Virial radius, $r_{\rm{200}} = r_{\rm{s}} c$.

Positions and velocities are assigned to the dark matter particles using the publically available algorithm {\it{mkhalo}} from the {\sc{nemo}} repository (\citealp{McMillan2007b}). Dark matter halos produced in this manner are evolved in isolation for 2.5 Gyr to test stability, and are found to be highly stable.

Our standard model has a dark matter halo mass of $10^{11}$M$_\odot$, consisting of 100,000 dark matter particles, with a concentration $c = 14$. The Virial radius is $r_{\rm{200}} = 95$~kpc. The peak circular velocity of the halo is 88~km~s$^{-1}$ at a radius of 15~kpc. If the halo were placed at a radius of 200~kpc within the smooth background potential of the cluster, the ratio of its central density is $\sim5$ times that of the surrounding ambient medium.

\subsubsection{The disk}
The stellar disk of the galaxy has an exponential form
\begin{equation}
\label{expdisk}
\Sigma(R) = \Sigma_0 {\rm{exp}} (R/R_{\rm{d}})
\end{equation}

\noindent
where $\Sigma$ is the surface density, $\Sigma_{\rm{0}}$ is central surface density, $R$ is radius within the disk, and $R_{\rm{d}}$ is the scalelength of the disk.

The scalelength and mass of the stellar disk is chosen to approximately match the observed properties of luminous Virgo dEs. These are characterised by close-to-exponential luminosity profiles with effective radii of the order of 1 kpc (\citealp{Janz2008}). Our fiducial model has a total stellar mass of $3.0 \times 10^{9}$M$_\odot$ ($3\%$ of the halo mass; \citealp{Peng2008}), and is formed from 20,000 star particles, distributed with $R_{\rm{d}}=0.75~$kpc, resulting in an effective radius of $\sim$1.25~kpc.
.
\begin{figure}
  \centering \epsfxsize=8.5cm
  \epsffile{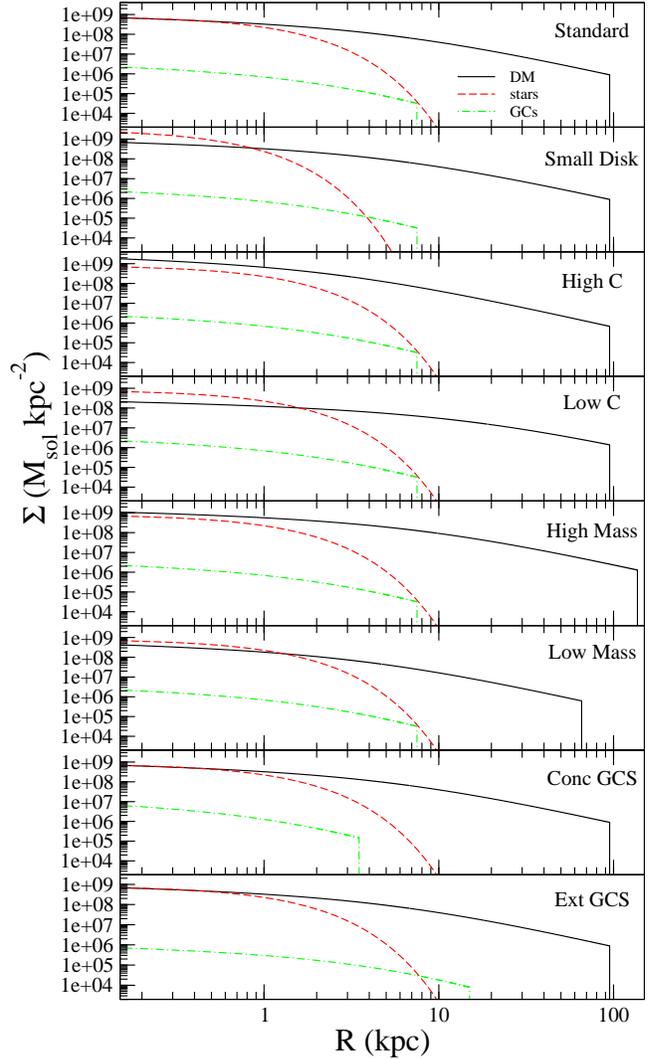}
  \caption{Surface density profiles of the dark matter (solid line), stars (dashed line), and GCs (dot-dashed line), for all the galaxy models.}
\label{sigprofiles}
\end{figure}

A radially varying velocity dispersion is chosen that ensures the disk is Toomre stable (\citealp{Toomre1964}) at all radii. In practice, a Toomre parameter of $Q>1.5$ is required throughout the stellar disk to ensure stability. We create a dispersion dominated disk with $v_{\rm{rot}}/v_{\rm{disp}}\sim0.25$. This results in a thick disk with axial ratio b/a $\sim0.6$ in agreement with the observed axial ratios of dwarf galaxies of this luminosity (\citealp{Lisker2007}; \citealp{Ruben2010}). A full description of the procedures followed to set up the galaxy disk can be found in \cite{Smith2010a}.

\subsubsection{The globular cluster distribution}
Each globular cluster is treated as a single N-body particle. The standard model has 60 equal-mass globular cluster particles with a summed total mass of $1.8\times10^7$~M$_\odot$, corresponding to a specific stellar mass fraction of 0.6 per cent (\citealp{Peng2008}; \citealp{Georgiev2010}). We model the globular cluster particle distribution as a Hernquist sphere: 

\begin{equation}
\rho(r) = \frac{M_{\rm{h}}}{2\pi}\frac{r_{\rm{h}}}{r}\frac{1}{(r+r_{\rm{h}})^3}
\label{herneqn}
\end{equation}
where $M_{\rm{h}}$ is the total mass of the Hernquist sphere, $r_{\rm{h}}$ is the Hernquist scalelength, and $r$ is radius. 

The  distribution is radially truncated at a cut-off radius of $7.5~$kpc. The observed GCS density distribution in luminous dEs depends on projected radius roughly as R$^{-1}$ (\citealp{Beasley2006}; \citealp{Puzia2004}). We obtain an adequate match to this radial dependency by choosing the Hernquist scalelength $r_{\rm{h}}=3.75~$kpc (half the cut-off radius). Particle velocities are assigned using the Jeans equation for an isotropic dispersion supported system.

Finally, we note that all models (combined dark matter halo, stellar disk and surrounding GCs) are evolved in isolation for 2.5~Gyr to ensure stability, before introduction into the cluster environment.

\subsection{A parameter study}
Our standard model is meant to roughly approximate the subclass of luminous dEs -- the only ones that have had their GCSs studied in sufficient detail. However, we are interested in understanding how the effects of harassment on any galaxy are dependent on the parameters of that particular galaxy model. We systematically vary one parameter at a time creating a set of 8 galaxy models as listed below.
\newline
\noindent
{\it{Halo concentration:}} We test a high concentration halo model (`High c') with c=30, and a low concentration halo model (`Low c') with c=5. The standard model has c=14.
\newline
\noindent
{\it{Halo mass:}} We test a high mass halo model (`High Mass') with M$_{200}=3.0\times10^{11}$M$_\odot$, and a low mass halo model (`Low Mass') with M$_{200}=0.333\times10^{11}$M$_\odot$. The standard model has M$_{200}=1.0\times10^{11}$M$_\odot$. Measured within 7.5 kpc (6 effective radii), the dynamical mass-to-light ratio for the high mass, standard, and low mass halo model is ${M}_{dyn}/{M}_{\star} \backsimeq$ 9, 5, and 4 respectively.
\newline
\noindent
{\it{Globular cluster distribution:}} We test a concentrated GCS (`Conc GCS') with Hernquist scalelength  $r_{\rm{h}}=1.875$~kpc, truncated at 3.75~kpc, and an extended GCS (`Ext GCS') with  $r_{\rm{h}}=7.5~$kpc, truncated at 15~kpc. For comparison, the standard model has  $r_{\rm{h}}=3.75~$kpc, and is truncated at 7.5~kpc.
\newline
\noindent
{\it{Disk scalelength:}} We test a small stellar disk model (`Small Disk') with exponential scalelength  $r_{\rm{d}}=0.375$~kpc, whereas the standard model has $r_{\rm{d}}=0.75$~kpc.
The surface density profile of each model is shown in Figure \ref{sigprofiles}.

We note that we do not attempt to simulate fainter cluster dEs or dSphs. The strong dependence of GC specific frequency on stellar mass for these galaxies (\citealp{Miller2007}; \citealp{Peng2008}) imply that essentially all dwarfs having $M_{\star} < 3\times10^{8}$M$_\odot$ host less than 10 metal-poor GCs. We therefore expect that the range of parameter space we cover should encompass the properties of most early type dwarfs who contain sufficient numbers of GCs to be useful for an enclosed mass measurement.

\begin{table}
\setlength{\tabcolsep}{2pt}
\centering
\begin{tabular}{|c|c|c|c|c|c|c|c|c|}
\hline
 Orbit & Orbit & r$_{\rm{init}}$ & r$_{\rm{min}}$ & r$_{\rm{max}}$ & N$_{\rm{peri}}$ & f$_{DM}$ & f$_\star$ & f$_{\rm{GC}}$\\
number & type & (kpc) & (kpc) & (kpc) & & \multicolumn{3}{|c|}{(standard)} \\
\hline
P1& Plunge & 987 & 287 & 973  & 1 & 0.16 & 0.89 & 0.98\\
P2& Plunge & 987 & 201 & 1250 & 1 & 0.02 & 0.48 & 0.38\\
P3& Plunge & 987 & 265 & 1024 & 1 & 0.23 & 0.97 & 1.00\\
P4& Plunge & 987 & 235 & 1793 & 1 & 0.18 & 1.00 & 0.98\\
P5& Plunge & 987 & 309 & 973 & 1 & 0.23 & 0.98 & 1.00 \\
P6& Plunge & 987 & 313 & 973 & 1 & 0.35 & 0.98 & 1.00 \\
P7& Plunge & 987 & 318 & 972 & 1 & 0.01 & 0.22 & 0.20 \\
P8& Plunge & 987 & 200 & 1953 & 1 & 0.13 & 1.00 & 0.97 \\
\hline
C1& Circular & 200 & 102 & 332 & 3 & 0.00 & 0.20 & 0.20 \\
C2& Circular & 200 & 111 & 401 & 3 & 0.04 & 0.68 & 0.70 \\
C3& Circular & 200 & 110 & 197 & 3 & 0.02 & 0.57 & 0.47 \\
C4& Circular & 200 & 124 & 201 & 3 & 0.00 & 0.00 & 0.00 \\
C5& Circular & 200 &  52 & 334 & 3 & 0.00 & 0.00 & 0.00 \\
C6& Circular & 200 & 124 & 240 & 3 & 0.08 & 0.86 & 0.92 \\
C7& Circular & 200 &  92 & 232 & 4 & 0.00 & 0.00 & 0.00 \\
C8& Circular & 200 & 107 & 292 & 3 & 0.01 & 0.30 & 0.23 \\
\hline
\end{tabular}
\caption{Orbit parameters. Columns (from left to right): (1) Orbit label, (2) Orbit type (plunging or circular), (3) Initial radius r$_{\rm{init}}$, (4) Minimum radius in cluster r$_{\rm{min}}$, (5) Maximum radius in cluster r$_{\rm{max}}$, (6) Number of pericentre passes N$_{\rm{peri}}$ over 2.5 Gyr, bound mass fraction of the standard model after 2.5~Gyr of harassment for the (7) dark matter f$_{DM}$, (8) stars f$_\star$, and (9) globular clusters f$_{\rm{GC}}$}
\label{orbittable}
\end{table}

\subsection{Orbits}
Our harassment model enables us to evolve a chosen galaxy model along a fixed orbit through the cluster. Each galaxy model then experiences an identical sequence of tidal encounters. Thereby we can fairly  compare the responses of different galaxy models to identical harassment histories. If the galaxy takes an alternative trajectory through the cluster, harassment histories are highly stochastic resulting in widely differing numbers and strengths of tidal encounters along the orbit. 

Although stochastic from orbit to orbit, stronger harassment typically occurs for orbits that spend  more time near the cluster centre where the number-density of harasser galaxies is highest, and the tidal force from the smooth background potential is largest (\citealp{Smith2010a}). In order to sample a range of harassment histories from strong to weak, we conduct 16 different orbits. The parameters of these orbits are shown in Table \ref{orbittable}. The first eight orbits (`orbit P1-P8') are on a plunging orbit with an initial radius of $\sim987~$kpc. The initial velocity is 450~km~s$^{-1}$ and purely tangential. In the absence of harassers, orbit P1-P8 would have a pericentre of $\sim230~$kpc. The latter eight orbits (`orbit C1-C8') have an initial radius of only 200~kpc from the cluster centre. The initial velocity is 1500~km~s$^{-1}$ and purely tangential. In the absence of harassers, orbits C1-C8 would be circular. However, harassment causes all the orbits to deviate, resulting in variation in the minimum and maximum radius from the cluster centre (see columns labelled `r$_{\rm{min}}$' and `r$_{\rm{max}}$').

The column labelled `N$_{\rm{peri}}$' indicates the number of times the galaxy passes pericentre along its orbit. The final three columns (labelled $f_{\rm{DM}}$, $f_\star$, and $f_{\rm{GC}}$) are the final bound fraction of the dark matter, stars and globular clusters, respectively, for the standard model. N$_{\rm{peri}}$, $f_{\rm{DM}}$, $f_\star$ and $f_{\rm{GC}}$ are measured after 2.5~Gyr of harassment along that orbit. $f_{\rm{DM}}$ can be considered an indication of the strength of the harassment along that orbit (i.e. a low $f_{\rm{DM}}$ indicates large mass loss of dark matter through strong harassment). Orbit P1-P8 typically experience weak harassment, as they make only a single plunging pass of their pericentre, and so spend little time near the cluster centre where the highest number density of harassers can be found. However strong mass loss can still occur in chance strong tidal encounters (e.g. see orbit P2). Orbits C1-C8 typically experience strong harassment as they remain close to the cluster centre throughout their orbit. In total, we conduct a total of 16 orbits on 8 galaxy models, resulting in a total of 128 harassment simulations.

\subsection{Measuring bound fractions}
The following technique is found to be a reliable, and reproducable method for measuring the fraction of particles that remain bound to the galaxy at any instant. In the first step, all particles within a 2.5~kpc radius of the centre of density of the galaxy are selected. In practice, the final bound fractions are found to be very insensitive to this choice of radius. Each particle confirms if it is bound to the other particles within this radius. Those that are bound to each other are considered a bound core within the galaxy center. Now an iterative procedure begins. In each iteration, all particles are tested to see if they are bound to the bound core. If they are, then their mass is added to the bound core. We call this method of growing the mass of the bound core the `snowballing' method. Then the iteration is repeated until the bound mass of the galaxy increases by no more than $1\%$ between iterations. In practice the total bound mass is found within 5-10 iterations. We find the snowballing method of measuring bound masses to be robust and trustworthy. A full description and testing of this approach will be presented in a future paper (\citealp{Farias2012}; in prep.)

\section{Results}

\begin{figure}
  \centering \epsfxsize=8.5cm
  \epsffile{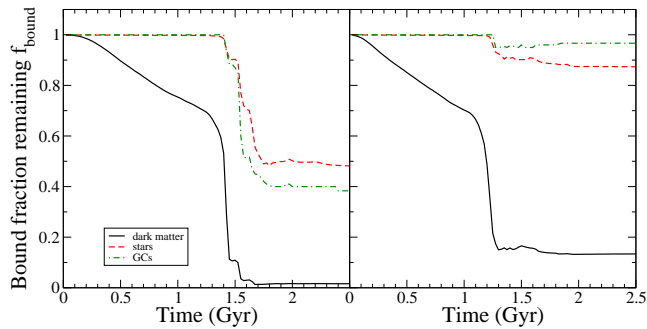}
  \caption{Plots of the time evolution of the fraction of bound particles of dark matter (black solid line), stars (red dashed line), and GCs (green dash-dotted line) for the standard galaxy model, undergoing harassment. The left panel is orbit P2 and is a medium strength harassment case with roughly half the stars and GCs stripped. The right panel is orbit P8 which is a weak harassment case, with $\sim10\%$ of stars and GCs lost}
\label{repegsfig}
\end{figure}

\begin{figure*}
  \centering \epsfxsize=18.0cm
  \epsffile{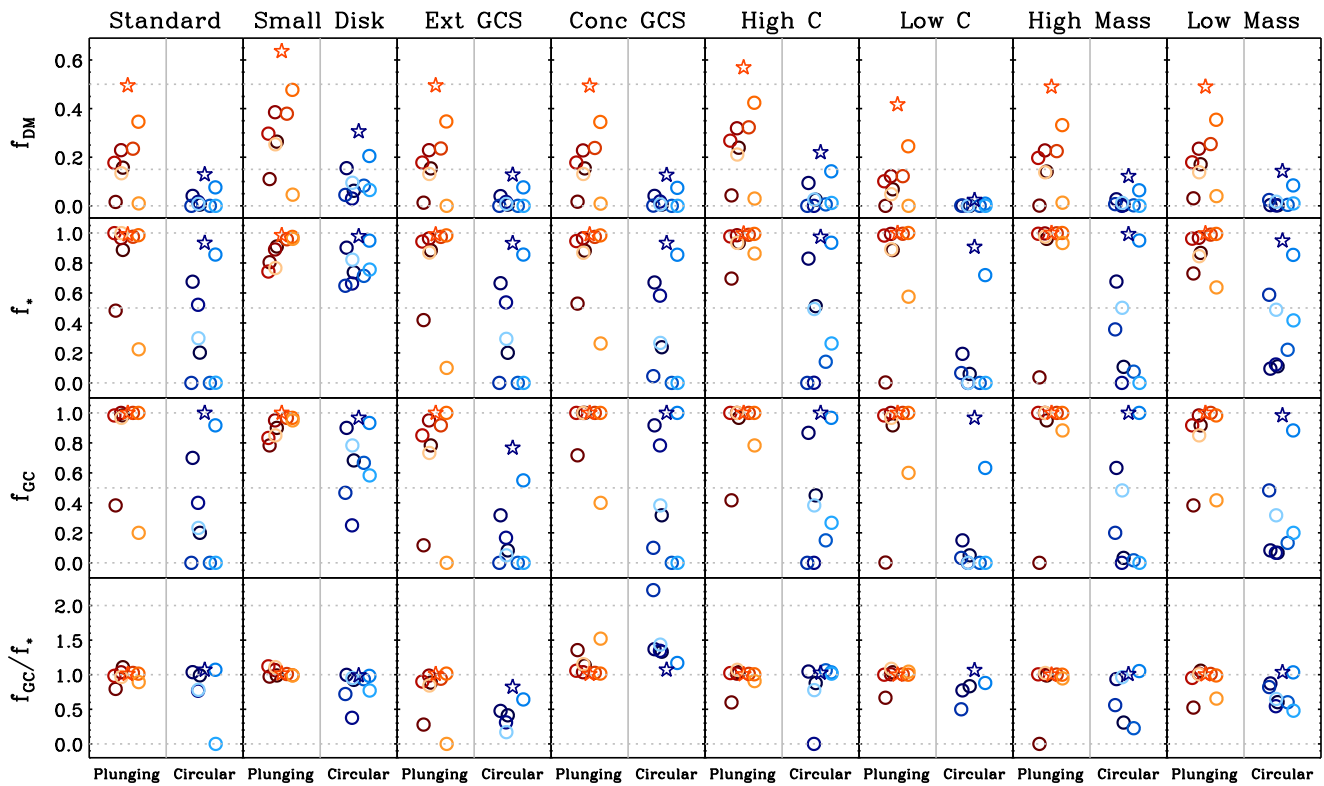}
  \caption{In the first, second, and third row we plot the bound fraction, measured at 2.5~Gyr of harassment, of dark matter ($f_{\rm{DM}}$), stars ($f_{\star}$), and globular clusters ($f_{\rm{GC}}$) respectively, against the orbit type; plunging (orange-red colours) or circular (light-dark blue colours). In the final row, the ratio of $f_{\rm{GC}}$/$f_{\star}$ is plotted indicating if stars or globular clusters have been preferentially lost. Columns indicate the model galaxy, as indicated by the column header. A particular orbit (e.g. orbit C1) has the same colour symbol in every panel. Circular symbols are results from the harassment simulations. A star symbol indicates a no-harasser orbit, where mass loss is purely from the tidal forces of the smooth background potential of the cluster. Please see Section \ref{masslossgalprops} for a detailed discussion.}
\label{masslossfig}
\end{figure*}

\subsection{Representative examples}
In Figure \ref{repegsfig}, we show two examples of mass loss of dark matter, stars and GCs, as a result of harassment acting on the standard model galaxy. The left hand panel is orbit P2 - an intermediate strength harassment case. The right hand panel is orbit P8, which is a case of weak harassment. However in both cases, a substantial fraction ($>80\%$) of the dark matter is unbound. This occurs rather gradually under the influence of the smooth, background potential of the cluster (from mass not associated with individual galaxies). However it can also occur rather rapidly when a high speed encounter with another cluster galaxy occurs (e.g. see the left panel at t=1.4~Gyr). In both the left and right panel, we note that the stars and globular clusters do not begin to be stripped until the dark matter fraction is low ($\sim$10-20$\%$). We shall better quantify this value in section \ref{whenstarslost}.

In all our simulations, we treat each individual GC as a point mass, whereas in reality an individual GC can be subject to tidally induced mass loss (e.g. \citealp{Baumgardt2007}; \citealp{Gieles2011b}; \citealp{Kupper2012}). Assuming that each of our GC contains 10$^5$~M$_\odot$ within a radius of 3~pc (equivalent to NGC104 or similar), we can calculate the Roche limit of a GC. The Roche limit is the maximum tidal force the GC can experience without being disrupted. We then monitor the tidal forces that each GC experiences during the orbit C1 infall. Orbit C1 is a case of strong harassment. In this simulation strong harassment results in heavy stripping of dark matter ($f_{\rm{DM}}=0.00$), stars ($f_{\star}=0.20$), and GCs ($f_{\rm{GC}}=0.20$). However, we find that GCs do not experience tidal forces greater than 0.1$\%$ of their own Roche limit, over the total 2.5~Gyr of harassment. Although many are stripped from the their host galaxy, they would not be completely disrupted by the tidal forces. Mass loss from real GCs also occurs due to internal processes operating within the GC, for example two-body encounters and stellar evolution (\citealp{gieles2012}). However, it is unlikely that these processes would result in disruption of many GCs during the short duration of the harassment simulations (2.5~Gyr). As our simulations use softened gravity between particles, we suppress the effects of dynamical friction acting on our GCs, which could result in their sinking towards the galaxy centre (\citealp{Oh2000}). However we expect the effect to be negligible for our model galaxy (i.e. our higher mass galaxy, with lower mass GCs results in a dynamical friction drag force that is $\sim$2-3 orders of magnitudes weaker than in the \citealp{Oh2000} study).

\subsection{Mass-loss dependency on galaxy properties}
\label{masslossgalprops}
We quantify the harassment induced mass loss that occurs to the dark matter, stars and GCs of all 8 galaxy models, for the 16 orbits we consider. To do this, we measure the fraction of dark matter ($f_{\rm{DM}}$), stars ($f_{\star}$) and globular clusters ($f_{\rm{GC}}$) that remain bound to the galaxy, after 2.5~Gyr in the cluster environment. Although we only present the bound mass fractions at 2.5~Gyr, we find the relative mass loss between models at earlier times (e.g. at 2~Gyr) is fairly represented. A low bound fraction therefore reflects that lots of mass has been stripped. We additionally conduct simulations where all the harasser galaxies are removed, and mass loss occurs purely as a result of the smooth backrgound potential of the cluster. We refer to these simulations as `no-harasser' runs, and conduct them for all model galaxies, on the plunging and circular orbits.  The results are shown in Figure \ref{masslossfig}.

In the first, second, and third row we plot $f_{\rm{DM}}$, $f_{\star}$, and $f_{\rm{GC}}$ respectively, against the orbit type (plunging or circular). In the bottom panel the ratio of $f_{\rm{GC}}$/$f_{\star}$ is plotted - this ratio can indicate if stars or globular clusters have been preferentially lost, and is directly proportional to the GC mass specific frequency, $T_{N} = N_{gc}/(M_{\star}/10^{9}$\,M$_{\odot})$ ({\citealp{Zepf1993}}). Columns indicate the model galaxy, as indicated by the column header. Plunging orbits are indicated by symbols with orange-to-red colours. Circular orbits are indicated by symbols with light-to-dark blue colours. A particular orbit (e.g. orbit C1) has the same colour symbol in every panel. Circular symbols are results from the harassment simulations. A star symbol indicates a no-harasser orbit, where mass loss is purely from the tidal forces of the smooth background potential of the cluster.

First, comparing the upper row to second row, it is clear that dark matter is always preferentially stripped in comparison to stars. The dark matter is more extended and so more easily unbound. However if we had instead plotted the number of dark matter particles in the same volume that encloses the stars, we would again see preferential loss of dark matter (as shown in \citealp{Smith2010a}). This indicates that the orbits of the dark matter particles are more plunging and eccentric than the stars - even if the halo is tidally truncated far beyond the stars, dark matter amongst the stars can still be effected. Our circular orbits (orbits C1-C8) tend to lose more dark matter than our plunging orbits (orbits P1-P8). However the stochastic nature of harassment leads to some (especially violent) plunging orbits losing more dark matter than some (especially gentle) circular orbits. For any specific orbit, the amount of dark matter is fairly insensitive to the galaxy model. This can be seen by the very similar distribution of points along the top row. For example, increasing halo concentration from c=5 to c=30 can reduce dark matter losses by a modest $\sim15\%$. The `small disk' model suffers the least dark matter losses, due to the tight gravitational binding between stars and local dark matter.

Now comparing top row to second and third row, it can be seen that baryonic losses only become significant once $f_{\rm{DM}}\sim0.15$ (as indicated by a horizontal dashed line in the top row). We shall better quantify the threshold dark matter fraction when baryons begin to be stripped in Section \ref{whenstarslost}. However as a result, our plunging orbits typically do not result in significant baryonic mass losses -- hence the clustering of points about high bound fractions. Two exceptions are the same two (especially violent) plunging orbits seen in the top row, where baryonic losses are roughly equal to the circular orbits. There is clearly large scatter in the baryonic mass loss for circular orbits. This indicates that once $f_{\rm{DM}}<0.15$ and the baryons begin to be stripped, the amount of baryons that are lost is very stochastic -- they are not sensitive to galaxy model, and depend more on the detailed interaction history. The one exception is the small disk model which is generally more robust to the effects of harassment in both dark matter and baryonic losses.

The fourth row indicates that in almost every case GCs and stars are lost in roughly equal amounts for most models ($f_{\rm{GC}}$/$f_{\star}\sim1$) when harassment is weak (for most of the plunging orbits). However when harassment is stronger, GCs are preferentially lost. This is related to their relative concentration -- half the mass of stars can be found within r$\sim$1.3~kpc, whereas half the mass of GCs are found within r$\sim$2.9~kpc. When the GCS is extended beyond the stars (the `ext GCS' model), the GCs are preferentially lost to an even greater extent. Only when the GCS is very concentrated (the `conc GCS' model), are stars preferentially lost. This is a natural expectation in the framework of the impulse approximation (\citealp{Spitzer1958}; \citealp{Gnedin1999}; \citealp{Gonzalez2005}).

We now consider the `no-harasser' simulations, where harasser galaxies have been removed from the cluster model. Except for the removal of the harasser galaxies, these are the same plunging and circular orbit as previously used in the harassment simulation. However, now mass loss is purely from the tidal forces of the smooth background potential of the cluster. It can be clearly seen that the star symbol (indicating a no-harasser simulation) appears above all the circular symbols (indicating the harassment simulations) in the upper, second and third row of Figure \ref{masslossfig}. This indicates that in all cases, the total mass loss from the smooth potential alone sets a lower limit on the mass loss when harassers are included. Therefore the addition of the harasser galaxies provides a source of additional mass loss beyond that of the background potential alone. This additional mass loss occurs from high speed tidal encounters, and the quantity of additional mass lost is very stochastic as it depends on the number of close encounters along that particular orbit. This is valid to first order, but we add that the inclusion of harasser galaxies does also modify the orbit of the model galaxy. In this way, harassers also cause second order changes in the mass loss from the background potential as well.

In summary, we have covered a considerable range of luminous early-type dwarf properties. These models should encompass the properties of the majority of the luminous early type dwarf galaxies, which contain sufficient numbers of GCs to be useful for enclosed mass measurements. We subject them to a wide range of strength of harassment. We find that their mass loss is rather insensitive to the parameters we have considered. Baryonic mass loss begins only when the bound dark matter fraction is low for all models. Once baryonic mass loss begins, the amount that is lost is very stochastic, and does not seem to depend sensitively on the galaxy parameters. The one exception is the model with the small disk, who is more robust to harassment in general. For all models, GCs are tidally stripped preferentially over stars, with the single exception of the model with a very concentrated spatial distribution of GCs.

\begin{figure}
  \centering \epsfxsize=9.0cm
  \epsffile{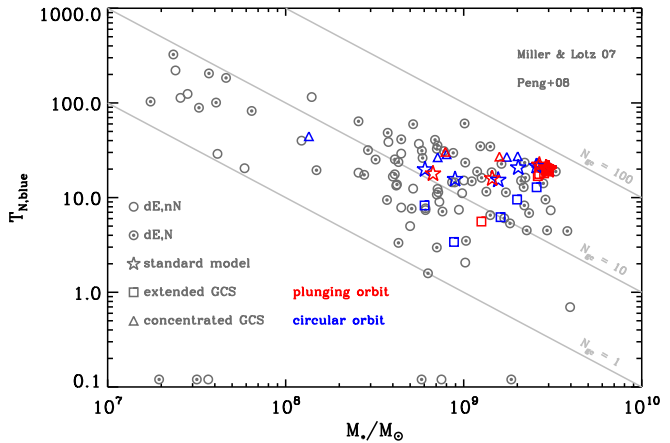}
  \caption{The metal-poor GC mass specific frequency ($T_{N,blue}$) versus stellar mass relation of actual (grey circles) and simulated Virgo dEs. The initial model consists of a $M_{\star} = 3\times10^{10}$ M$_{\odot}$ dE with $T_{N,blue} = 20$. Different galaxy models and orbits are coded with different symbols and colours, respectively. For any given galaxy model, circular orbits (blue) preferentially result in much larger mass losses than the corresponding plunging orbits (red). The initial spatial distribution of the GCS also plays a relevant role, as expected from the impulse approximation. Thus, models with more extended GCSs (squares) tend to preferentially lose GCs with respect to stars -- therefore decreasing their original $T_{N,blue}$. The opposite is true for models with more concentrated GCSs (triangles), while the standard models (stars) lose both components with a similar efficiency.}
\label{rubenfig}
\end{figure}

\begin{figure}
  \centering \epsfxsize=8.5cm
  \epsffile{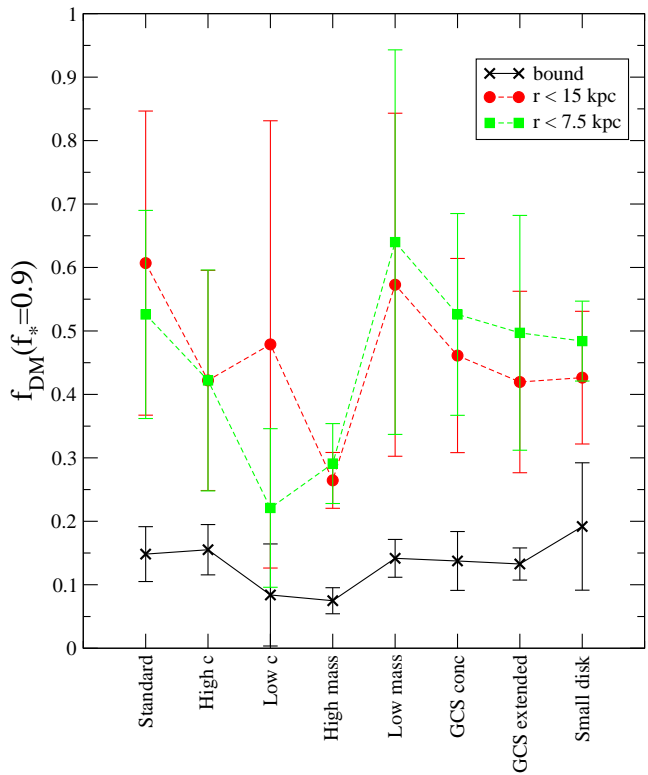}
  \caption{Fraction of bound dark matter particles remaining when the bound fraction  of stars reaches 0.9 (when $10\%$ of the stars have been unbound) for each galaxy model. Solid line \& cross symbols (black) is the bound dark matter fraction. Dashed lines are fractions of the pre-harassment density of the dark matter measured within a spherical volume of radius 15 kpc (circle symbols; red) or radius 7.5 kpc (square symbols; green). Significant loss of stars occurs when the bound dark matter fraction reaches about $10-15\%$ and this value is fairly model independent. This is when the dark matter density surrounding the globular clusters has roughly halved.}
\label{fcritdmfig}
\end{figure}

\subsection{Specific frequencies: Comparison with observations}
In Figure \ref{rubenfig} we present a direct comparison between observations and three of our models at the end of the simulation. Grey circles show the relation between the metal-poor GC mass specific frequencies ($T_{N,blue}$) of actual Virgo dEs (from Peng et al. 2008 and Miller \& Lotz 2007) and the host galaxy stellar mass. Our initial model has $M_{\star} = 3\times10^{9}$ M$_{\odot}$ and 60 GCs -- all assumed to be metal-poor for the sake of this comparison (\citealp{Peng2006}) -- and therefore $T_{N,blue} = 20$. Different models and orbits are coded with different symbols and colours, respectively. Thus, plunging orbits are represented in red, while blue symbols correspond to circular orbits. Stars, squares and triangles indicate the final stages of the standard, the `extended GCS' and the `concentrated GCS' models, respectively. It is obvious that both the internal configuration of galaxy components and its orbit within the cluster play a relevant role. Our circular orbits result in the most significant mass losses, both for stars and GCs. Our plunging orbits inflict less damage (also see \citealp{Smith2010a}), and it is tempting to suggest they are more probable for actual dEs harbouring rich GCSs (we shall study this further in a follow-up investigation). However, the initial GCS spatial distribution also influences the final location of the remnant in this diagram. Models having more extended GCSs (squares) tend to lose a larger fraction of GCs than stars, and therefore their $T_{N,blue}$ systematically decreases from its initial value (c.f. Figure \ref{masslossfig}). The standard model (stars) shows an almost flat behaviour, indicating that both GCs and stars are lost with a similar efficiency. Finally, only models with concentrated GCSs (triangles) exhibit a $T_{N,blue}$ increase, and this is always relatively moderate -- a recent, harassment-induced origin from more massive progenitors appears to be unfeasible for the lowest-mass, highest-$T_{N,blue}$ Virgo dEs. These results are consistent with simple theoretical considerations of interactions in the impulsive regime (see \citealp{Ruben2012} for a detailed discussion).

\subsection{The amount of dark matter remaining when baryons begin to be stripped}
\label{whenstarslost}
We re-emphasise that the stars begin to be stripped when the bound dark matter fraction $f_{\rm{DM}}$ becomes low. To better quantify the required value of $f_{\rm{DM}}$, we record the value of $f_{\rm{DM}}$ when $f_{\star}$ reaches 0.9. In other words, when $10\%$ of the stars have been stripped, we note the bound dark matter fraction. This is denoted $f_{\rm{DM}}(f_{\star}=0.9)$, and is plotted on the y-axis of Figure \ref{fcritdmfig} (black crosses). Along the x-axis is the galaxy model. For every model galaxy, we obtain the average and standard deviation of the value of $f_{\rm{DM}}(f_{\star}=0.9)$ for all 16 orbits. Despite a wide range in strength of harassment, and galaxy model properties, the $f_{\rm{DM}}(f_{\star}=0.9)$ of bound dark matter particles (cross symbols; black) is roughly constant with a value between 0.15-0.20. In other words, {\it{once the remaining bound dark matter fraction has fallen below 0.15-0.20, significant stellar stripping begins}}. This result can be seen, by eye, for the two orbits shown in Figure \ref{repegsfig}, but now we see it remains valid for a range of galaxy models and over a much larger variety of orbits. Similar results are seen in simulations of dwarf galaxies undergoing mass loss as they orbit within the Milky Way potential (\citealp{Penarrubia2008}), despite the fact that mass loss is often driven by high speed galaxy-galaxy encounters in our models.

We also measure the total number of dark matter particles within spheres of radius 15~kpc and 7.5~kpc, centred on the galaxy centre, when 10$\%$ of the stars have been stripped. We normalise these values by their initial value, prior to harassment, and plot these in Figure \ref{fcritdmfig} as well (red circles, green squares). When 10$\%$ of the stars had been stripped, typically the dark matter density surrounding the stellar disk and GCs had fallen by roughly a factor of 2, although the value is quite scattered between models. This explains why stars and GCs remain mildly affected, even when $\sim80\%$ of the total dark matter has been unbound - the  density of dark matter surrounding the stars and GCs has only halved.

In summary, stars and GCs begin to be stripped in significant quantities ($>10\%$ stripped) once the total bound dark matter fraction $f_{\rm{DM}}$ falls below 10-15$\%$. This is true for all models, and so is insensitive to the the fairly large range of galaxy parameters we have considered. In the following section, we investigate to what extent the harassment and stripping of GCs impacts their dynamical state.

\begin{figure}
  \centering \epsfxsize=8.5cm
  \epsffile{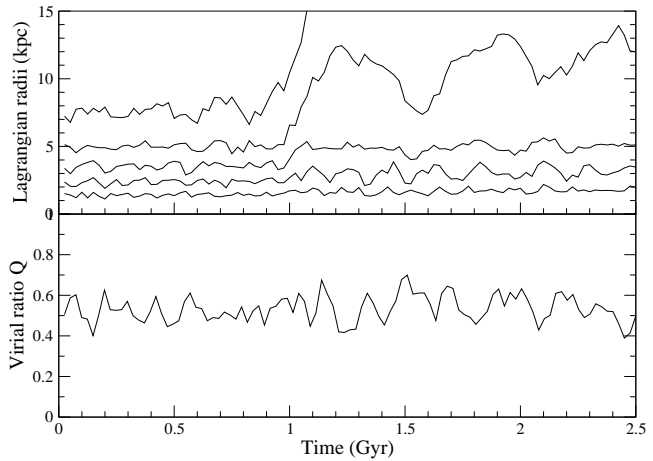}
  \caption{Upper panel: The evolution of the 20, 40, 60, 80, 95$\%$ Lagrangian radii of the GCs, for the standard model on the orbit P4 trajectory. While the inner most GCs are unaffected, some GCs are stripped, while some take up a more extended configuration. This is mainly in response to the reduced potential well in which the GCs orbit as mass is stripped from the galaxy. Lower panel: The evolution of the virial ratio of the bound GCs shows no significant deviation from virialised when outer GCs are stripped, and demonstrates the new more extended configuration of GCs quickly returns to a dynamically relaxed state within the new potential well surrounding the GCs.}
\label{ellipharass4fig}
\end{figure}

\subsection{The dynamical state of the globular cluster system (GCS)}

\subsubsection{Evolution of the spatial distribution of the GCS}
The GCS of late-type dwarfs typically do not extend beyond $\sim$2 effective radii (\citealp{Georgiev2009}). Yet in dEs the GCSs appear to extend to 4-7 galaxy effective radii (\citealp{Beasley2006}; \citealp{Beasley2009}). Here we test if high speed tidal encounters could result in a more extended spatial distribution of the GCs that remain bound to a dwarf galaxy.

In Figure \ref{ellipharass4fig} we plot the time evolution of the Lagrangian radii, and virial ratio Q\footnote{The virial ratio is defined as the absolute value of the total kinetic energy of the GCs divided by the total potential energy of the GCs. Thus a virialised system has Q=0.5.}, of the GCS of the standard model, for orbit P4. This trajectory results in fairly mild harassment with mild star and GC losses ($<10\%$), but still substantial dark matter losses ($\sim75\%$). The upper panel shows the 20, 40, 60, 80, \& 95$\%$ Lagrangian radii of the GCs (from bottom to top respectively). The innermost GCs are unaffected even when the outer GCs are stripped. However some of the GCs, {\it{that are still bound}}, take up a new, more extended distribution that appears to be stable with time. This is primarily in response to the reduced potential well in which the GCs orbit. The potential well is reduced by harassment induced mass loss. The lower panel shows the evolving virial ratio of the bound GCs. This reveals that the new extended GCS is indeed stable, and has quickly relaxed back into dynamical equilibrium with the more shallow potential well within which the GCs orbit.

\begin{figure}
  \centering \epsfxsize=8.5cm
  \epsffile{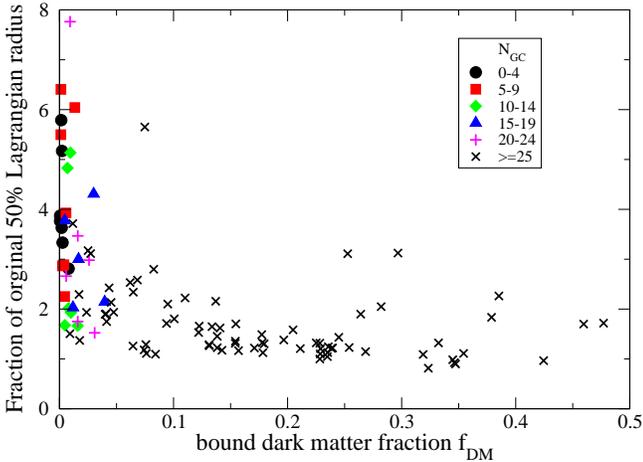}
  \caption{The fractional amount that a radius containing half the bound GCs increases by the end of the simulation. This is an indicator of how the bound GCS has expanded spatially, and is plotted against the bound dark matter fraction. Symbols indicate the number of bound GCs after harassment (as shown in the key). There is no clear difference in behaviour between galaxy models. Typically the spatial distribution of their GCSs experience no more than a factor of two increase, due to harassment. However as the galaxies approach complete disruption ($f_{\rm{DM}}<5\%$), the remaining bound GCs ($N_{\rm{GC}}<25$) may expand by as much as a factor of 4-8.}
\label{Qflagrradfig}
\end{figure}

To test the extent to which this `puffing-up' of the bound GCS can occur, we measure the fraction by which the $50\%$ Lagrangian radius of the bound GCs changes for the standard model after 2.5 Gyr of evolution, for all 16 orbits. For example, if the fraction is 2, then the bound GCs have roughly doubled their extension. Of course unbound GCs can be stripped and moved to much larger radii. However we emphasise that here we consider the change in the spatial distribution of the {\it{bound}} GCs, as these will remain in their new extended distribution. We plot this against the final bound dark matter fraction $f_{\rm{DM}}$ of all models, along all orbits, in Figure \ref{Qflagrradfig}. We find no clear trend with model galaxy, therefore symbols represent the final remaining number of bound GCs (as illustrated in the key). Although there is some scatter, as a rule of thumb most galaxies with $f_{\rm{DM}}>0.05$ expand by as much as a factor of 2-3. Only when the galaxy is much closer to complete disruption ($f_{\rm{DM}}<0.05$), can the bound GCs expand more substantially by factors as much as $\sim$4-8. We repeat all measurements using the $75\%$ Lagrangian radius of the bound GCs, instead of the $50\%$ Lagrangian radius. We find that on average the amount of expansion is only slightly greater (by a factor of 1.20$\pm$0.35) if we use the $75\%$ Lagrangian radius. We defer a detailed study of the link between orbit and spatial distribution to a future paper. 

In summary, due to harassment the current distribution of bound GCs about a cluster dE might be more extended than the original distribution by roughly a factor of 2-3 if the galaxy is not close to complete disruption ($f_{\rm{DM}}>0.05$).

\subsubsection{How close to virialised are harassed GCSs?}
There are many sources of uncertainty that may effect a measurement of the enclosed mass using GC dynamics. However one of the most fundamental of these is the assumption that the GCS is in dynamical equilibrium within the  potential of the galaxy. We first consider the virial state of bound members of the GCSs under the influence of harassment, before then considering other sources of error. We emphasise that we only consider the dynamics of the {\it{bound}} GCs in calculating the virial ratio. This is useful as how close the bound members of the GCS are to dynamical equilibrium sets a lower limit on the uncertainty in any enclosed mass estimate, with other sources of error only adding additional uncertainty.

\begin{figure}
  \centering \epsfxsize=8.5cm
  \epsffile{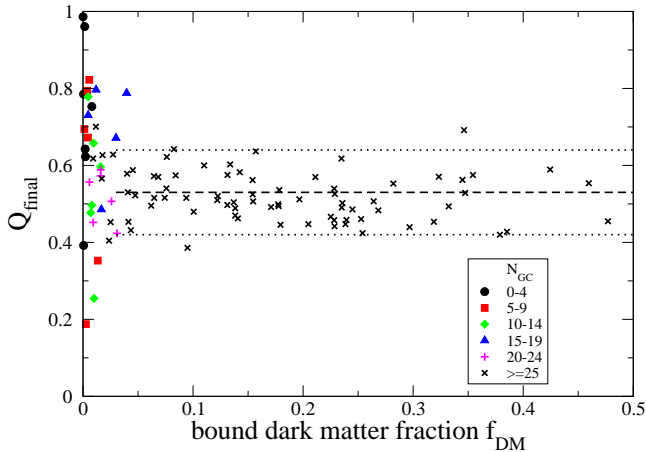}
  \caption{The virial state of the bound GCs of the GCS, at the end of the simulation as a function of the bound dark matter fraction. Symbols indicate the number of GCs that remain bound (indicated in the key). We see no obvious dependency on the galaxy model used. When the bound dark matter fraction is very low ($f_{\rm{DM}}<0.03$) and the galaxy is close to complete disruption, the virial state of the bound GCs may become super-virial (Q$>$0.6). In this regime, the assumption of virial equilibrium could result in errors of as much as a factor of 2. However, when $f_{\rm{DM}}>0.03$, the virial state is typically close to virialised: Q=0.53$\pm0.11$ (dashed line is mean, dotted lines are standard deviation). The upper-limit is Q=0.7. Therefore, if a virialised GCS is assumed, the total enclosed mass will be under- or over-estimated by (at most) $40\%$.}
\label{Qvirrealfig}
\end{figure}

In Figure \ref{Qvirrealfig}, the virial state of the bound GCs of each model, after 2.5~Gyr of harassment, is plotted against the final bound dark matter fraction $f_{\rm{DM}}$. Symbols represent the final number of bound GCs of that particular model (see key for reference). We include all model galaxies, for all 16 orbits with the exception of those who have no GCs remaining with which to measure the virial ratio. The trend has no clear dependency on galaxy model. 

On average the virial ratio is 0.53$\pm$0.11 for all models with $f_{\rm{DM}}>$0.03. The mean (dashed horizontal line) and standard deviation (dotted horizontal line) are shown on the plot. If $f_{\rm{DM}}>$0.03, the full range of virial ratio for our models is Q=0.4-0.7. For the upper-bound value of Q=0.7, the enclosed mass would be overestimated by at least $40\%$ if the GCS were assumed to be in dynamical equilibrium. To put it another way, {\it{mass estimates are - at most - $40\%$ overestimated by assuming virial equilibrium, even when a galaxy has lost $97\%$ of its dark matter through harassment, and dark matter density surrounding the GCs has fallen by more than a factor of two!}}

However, once $f_{\rm{DM}}$ falls below the critical value of $f_{\rm{DM}}$=0.03, the spread in virial ratio of the GCS grows (Q=0.19-0.99). If virial equilibrium is assumed in this regime, then enclosed mass estimates could be over- or under-estimated by a factor of $\sim2$ in the most extreme cases. Some of this increased scatter can be attributed to the low-number statistics of measuring the virial ratio with $<10$ bound GCs. However we note that there is assymetry in the distribution of Q values for $f_{\rm{DM}}<$0.03, with $57\%$ found at Q$>$0.6 compared to only $11\%$ found at Q$<$0.4. This would not result from simple low-number statistics, and indeed some of the Q$>$0.6 points have between 15-19 GCs. Therefore the tidal encounters of harassment can maintain the GCSs in a super-virial state (Q$>$0.6).

To further confirm that a boosted virial ratio is a genuine property of the GCS where the galaxy has $f_{\rm{DM}}<0.03$, and not just due to the low number of GCs remaining, we repeat three of the standard model harassment orbits, except we replace the 60 GCs of the standard model with 6000 GCs (but with the same total mass in GCs). We refer to this model as the `high N' model. We choose three orbits that result in $f_{\rm{DM}}<0.03$; orbit C1, C5 and C8 (see Table \ref{orbittable} for remaining bound fractions with the standard model).  Clearly this number of GCs is highly unrealistic for a galaxy of this luminosity, but the virial ratio of the `high N' model is free from low-number effects. Bound dark matter and stellar fractions are identical. As we have only changed the number of GCs, this would be expected. Bound GC fractions differ by no more than 10$\%$, and therefore are not sensitive to the number of GCs we initially assume. We calculate the average virial ratio of the standard and high N model over the 2.5~Gyr of harassment. Orbit C1 has Q=0.58$\pm$0.11 (standard model) and Q=0.60$\pm$0.07 (high N model). Orbit C5 has Q=0.65$\pm$0.08 (standard model) and Q=0.65$\pm$0.06 (high N model). Orbit C8 has Q=0.58$\pm$0.09 (standard model) and Q=0.59$\pm$0.11 (high N model). We see that the average Q is always boosted to be supervirial, independently of GC number, but the standard deviation of the standard model Q is greater due to low number effects. This underlines that when $f_{\rm{DM}}<0.03$, dynamically the GCSs are maintained supervirial by tidal encounters.

In summary, galaxies that have lost up to 97$\%$ of their dark matter can harbour a GCS whose virial state may cause enclosed mass estimates to be under- or over-estimated by no more than 40$\%$. Therefore deviations from virial equilibrium play only a minor role in the mass estimate uncertainties. However once $<3\%$ of the bound dark matter remains, the bound members of the GCS may become super-virial as the galaxy approaches complete disruption. In this $<3\%$ regime, the assumption of virial equilibrium could result in mass estimates being incorrect by as much as a factor of 2. It is difficult to see how the virial state of the GCS could be determined independently. There are also additional sources of error beyond the assumption of virial equilibrium, that we shall consider in the following section. Therefore these errors can be considered lower limits on the true uncertainties in an enclosed mass estimate.

\subsubsection{Additional errors affecting pointed observations}
To convert the dynamics of a GCS into an enclosed mass, and in the absence of rotation, 
an equation may be used of the following form:

\begin{equation}
M_{\rm{obs}}(<r_{\frac{1}{2}{\rm{GCS}}})=\eta r_{\frac{1}{2}{\rm{GCS}}} \sigma^2
\label{massenclosedeqn}
\end{equation}
\noindent
where $M_{\rm{obs}}(<r_{\frac{1}{2}{\rm{GCS}}})$ is the total mass enclosed within a projected radius containing half the clusters of the GCS ($r_{\frac{1}{2}{\rm{GCS}}}$), $\eta$ is a normalising constant that differs depending on the mass distribution, and $\sigma$ is the line-of-sight velocity dispersion of the GCs within $r_{\frac{1}{2}{\rm{GCS}}}$.

In Equation \ref{massenclosedeqn}, the GCS is assumed to be in dynamical equilibrium. As demonstrated in Figure \ref {Qvirrealfig}, the GCS may not be in dynamical equilibrium, causing a first source of error. Also the GC and mass distribution is assumed to be spherical, and the observed velocity dispersion may be assumed to be isotropic. In a pointed observation of the GCS of a galaxy orbiting within the cluster environment, a number of these assumptions may not be valid. We refer to these issues as `line-of-sight' effects - additional sources of error due to the fact that the galaxy is observed in projection. For example, the GC distribution may not be exactly spherical at any instant - even in isolation - due to their low numbers. Furthermore, along its orbit a galaxy may be stretched at pericentre and contracted at apocentre, by the tidal field of the cluster. This may further exasperate how non-spherical the GC distribution is, as well as how isotropic their velocity dispersion is. Finally, along the line-of-sight, high velocity unbound GCs may appear close to the body of the galaxy, and be included in the sample, potentially resulting in boosting of the velocity dispersion (and an overestimate of the enclosed mass). 

To use Equation \ref{massenclosedeqn}, we must first find $\eta$ which is density distribution dependent. For a simple mass distribution, $\eta$ can be calculated analytically. The mass distribution enclosed by the GCS is a combination of the inner NFW dark matter halo, and an exponential disk of stars. For simplicity, we numerically calculate $\eta$ for each of our galaxy models: While evolving our model galaxies in isolation, we measure $r_{\frac{1}{2}{\rm{GCS}}}$, the velocity dispersion of the GCS within $r_{\frac{1}{2}{\rm{GCS}}}$, and finally the total mass (in dark matter and stars) enclosed in $r_{\frac{1}{2}{\rm{GCS}}}$. We can then solve Equation \ref{massenclosedeqn} for $\eta$. For the standard model we find $\eta=(5.8 \pm 1.7) \times10^{5}$ M$_\odot~kpc^{-1}~(km~s^{-1})^{-2}$. We also calculate $\eta$ for the other galaxy models - it does not vary substantially between the models and the average value is $\eta=(5.7 \pm 1.7) \times10^{5}$ M$_\odot~kpc^{-1}~(km~s^{-1})^{-2}$ ($\pm30\%$ errors).  {\it{Therefore mean enclosed masses calculated, using Equation \ref{massenclosedeqn}, have $30\%$ errors from $\eta$ alone, as well as additional errors from `line-of-sight' effects and the assumption of dynamical equilibrium.}}

\begin{figure}
  \centering \epsfxsize=8.5cm
  \epsffile{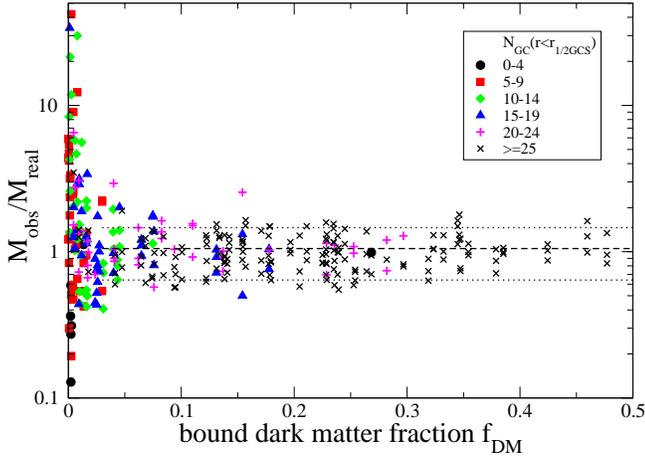}
  \caption{The observed mass over the real mass as a function of the bound dark matter fraction. The observed mass $M_{\rm{obs}}$ is calculated using Equation \ref{massenclosedeqn}, whereas $M_{\rm{real}}$ is the real mass measured directly from the simulation. If $f_{\rm{DM}}$ is greater than 3$\%$, the observed mass can be estimated to within a factor of two due to a combination of deviation from dynamical equilibrium, uncertainty in $\eta$ (see Equation \ref{massenclosedeqn}), and additional line-of-sight effects (such as unbound GCs). The dashed and dotted horizontal lines show the average and standard deviation ($M_{\rm{obs}}/M_{\rm{real}}$=1.05$\pm$0.41) for $f_{\rm{DM}}>3\%$. Symbols indicate the number of GCs found within $r_{\frac{1}{2}{\rm{GCS}}}$, and therefore used to calculate the observed mass. In some cases, the mass is measured to within a factor of two using only $<15$ GCs. However for $f_{\rm{DM}}<3\%$, when the galaxy is close to complete disruption, mass estimates can be substantially overestimated.}
\label{massobsfig}
\end{figure}

To quantify how much these combined effects may influence our mass estimates, we first use Equation \ref{massenclosedeqn} on all our model galaxies after 2.5~Gyr of harassment. For this, we use the average value found for $\eta$ ($\eta=5.7 \times10^{5}$ M$_\odot~kpc^{-1}~(km~s^{-1})^{-2}$) and, for each individual model, we choose three independent lines of sight (along the x, y, and z axis). This gives us 384 data points in total (3 lines of sight and 128 models). We exclude galaxies that are completely disrupted ($f_{\rm{DM}}=f_\star$= 0.00) leaving 330 data points in total. In order to mimic the observations, we assume a field-of-view that includes GCs up to 8 effective radii (10~kpc) from the galaxy. We later demonstrate our results are not sensitive to this rather arbitary choice of field-of-view. We then measure $r_{\frac{1}{2}{\rm{GCS}}}$ of the GCSs in the field-of-view, and the velocity dispersion of the GCs within $r_{\frac{1}{2}{\rm{GCS}}}$. Then we calculate the estimated enclosed mass for each line-of-sight ($M_{\rm{obs}}$) using Equation \ref{massenclosedeqn}. Finally we divide each value by the true enclosed mass within a radius equal to $r_{\frac{1}{2}{\rm{GCS}}}$ along that line of sight ($M_{\rm{real}}$), providing us with the quantity $M_{\rm{obs}}/M_{\rm{real}}$. This ratio directly tells us by what factor we measure the true enclosed mass, and is plotted on the y-axis of Figure \ref{massobsfig}. We do not differentiate between different galaxy models in Figure \ref{massobsfig}, as we find no obvious dependency on galaxy model. Instead we provide each data point with a symbol according to the number of GCs found within $r_{\frac{1}{2}{\rm{GCS}}}$ (and thus the number of GCs used to calculate the velocity dispersion for that data point).

Below the critical bound dark matter fraction $f_{\rm{DM}}=0.03$, the mass estimates typically become overestimated. Of all galaxies in this regime, 35$\%$ have their real mass over estimated by more than a factor of two. Only 17$\%$ recover the true mass to within 20$\%$. As shown in Figure \ref{Qvirrealfig}, the bound GCs are often supervirial when $f_{\rm{DM}}>0.03$. However this could only result in mass over estimates by a factor of two of the true mass at most. The key cause of such boosted mass estimates appears to be primarily unbound GCs, seen down the line of sight. With such a low dark matter fraction, the galaxy is close to complete disruption, and a large number of the previously bound GCs may suddenly find themselves unbound. Therefore a large number of unbound GCs are seen in projection to be close to the galaxy, and their velocity dispersion is large, resulting in boosted mass estimates.

For $f_{\rm{DM}}>0.03$, the observed mass recovers the true mass well for all data points, but with significant scatter - $M_{\rm{obs}}/M_{\rm{real}}$ is on average 1.05$\pm$0.41 (the horizontal dashed line is the mean, and the horizontal dotted lines are the standard deviation). The scatter is a little larger than the $\pm20\%$ standard deviation we get from the virial ratio alone, due to the additional line-of-sight errors and uncertainty in the normalising constant $\eta$ (see Equation \ref{massenclosedeqn}). However, taking into account the full range in $M_{\rm{obs}}/M_{\rm{real}}$ seen, {\it{above $f_{\rm{DM}}\sim3\%$, mass estimates are accurate to within a factor of 2}}. We note that some green diamond symbols can be found close to $M_{\rm{obs}}/M_{\rm{real}}=1$. Therefore, mass measurements that are accurate to within a factor of two can be made using the velocities of only 10-14 GCs.

We repeat all measurements using a field-of-view encompassing all GCs within 12 effective radii (15~kpc) of each model, and find our statistics are in close agreement to when our field-of-view was 8 effective radii (10~kpc). Thus the accuracy of our mass estimates are not sensitive to our rather arbitary choice of field-of-view. Unbound GCs do not significantly increase enclosed mass errors in the $f_{\rm{DM}}>0.03$ regime as once they are unbound, they quickly separate from the main body of the dwarf galaxy and few are seen along the line-of-sight. This remains the case even when we increase our field-of-view out to 12 effective radii. These stripped GCs can make an important contribution to the intracluster component, as seen recently in nearby clusters (\citealp{Lee2010}; \citealp{Peng2011}; \citealp{West2011}). We defer a study of the properties of the stripped GCs for a future paper.

In summary, as long as the remaining bound dark matter fraction $f_{\rm{DM}}>0.03$, and assuming a non-rotating GCS, an enclosed mass may be calculated using Equation \ref{massenclosedeqn} and $\eta=5.7 \times10^{5}$ M$_\odot~kpc^{-1}~(km~s^{-1})^{-2}$. This method of mass-measurement can be applied to galaxies with as few as 10-15 GCs. However non-virialised GCSs, line-of-sight effects, and uncertainty in the assumed $\eta$ provide sources of error. As a result, we find the calculated enclosed mass agrees with the true enclosed mass to within a factor of two if the bound dark matter fraction is above a critical value $f_{\rm{DM}}>0.03$.

\subsection{Determining if a galaxy is above the critical dark matter fraction}
Using GC dynamics to measure the total enclosed mass is clearly only reliable if a galaxy has more than the critical amount of bound dark matter ($f_{\rm{DM}}>0.03$). One possible means to test if this is the case is by measuring the dynamics of the stellar disk.

It is well established that, if harassment is strong, it can effectively heat fragile stellar disks and convert them from rotationally supported systems into dispersion supported systems. Therefore we test the hypothesis that the amount of rotation might provide constraints on the bound dark matter fraction.

\begin{figure}
  \centering \epsfxsize=8.5cm
  \epsffile{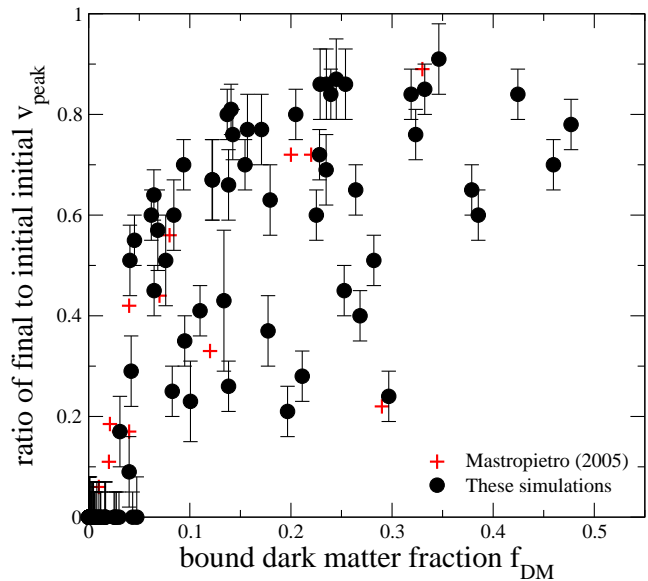}
  \caption{The ratio of the final to initial $v_{\rm{peak}}$ (the peak in the rotation curve) as a function of the dark matter fraction. Filled circular symbols (black) with error bars are the model galaxies. Plus symbols (red) are harassed dwarf galaxies from \citet{Mastropietro2005}, and represent lower limits. The peak value in the rotation curve of harassed dwarfs, $v_{\rm{peak}}$, decreases as dark matter is lost. As $f_{\rm{DM}}$ falls below 0.05, the peak rotation velocity falls rapidly. $v_{\rm{peak}}$ is typically reduced to 0-20$\%$ of its pre-harassment value as $f_{\rm{DM}}$ approaches zero.}
\label{vpeakfig}
\end{figure}

For all our model galaxies, falling along all orbits, we produce the rotation curve of the stellar disk after 2.5 Gyr of harassment, and record the peak value of rotation $v_{\rm{peak}}$. To do so, we choose our lines of sight to be along the plane of the disk, so as we do not have to correct for inclination. We then calculate the average line of sight velocity through the disk, in bins of distance from the disk centre. Our bins are 1 kpc wide, and we typically measure the rotation curve out to 10 kpc which is sufficient to encapsulate $v_{\rm{peak}}$. We reject bins with $<50$ particles to reduce noise in rotation curves at large distances from the disk centre. Although our galaxy disks are dispersion dominated from the beginning, this method clearly and reliably recovers the rotation curve. 

We note the $v_{\rm{peak}}$ with estimated errors for each orbit, and divide these by the initial $v_{\rm{peak}}$ before harassment. This gives us the fraction of the initial $v_{\rm{peak}}$ as a result of harassment. We plot this fraction against the remaining bound dark matter fraction $f_{\rm{DM}}$ in Figure \ref{vpeakfig}. Our simulations are shown as filled circles (black) with error bars. We also include data points from the simulations of \cite{Mastropietro2005} (red plus symbols), where we take $v_{\rm{peak}}$ values from their provided rotation curves. The initial $v_{\rm{peak}}$ of their model galaxy is not specifically given, so we assume $v_{\rm{peak}}=92$km~s$^{-1}$, based on their Fig. 1 in \cite{Mastropietro2005}. Also the inclination of their disks is unknown, therefore all of their data points can be considered lower limits.

The peak value in the rotation curve of harassed dwarfs, $v_{\rm{peak}}$, tends to decrease as dark matter is lost. At a fixed $f_{\rm{DM}}$, there is clearly some variation in the fraction of the initial $v_{\rm{peak}}$. However the upper-most value of the fraction falls rapidly when $f_{\rm{DM}}$ falls below about 0.1. $v_{\rm{peak}}$ is typically reduced to 0-20$\%$ of its pre-harassment value as $f_{\rm{DM}}$ approaches zero. In short, the amount of rotation in the stellar disk is strongly reduced when the remaining bound dark matter becomes small. 

We therefore suggest that rotation in the stellar disk could provide an indication of whether GCS dynamics are useful to estimate the enclosed mass. For example, consider a disk that is strongly supported by rotation. It is likely that such a galaxy still contains sufficient bound dark matter, that its GCS dynamics are useful to estimate the enclosed mass. Otherwise the disk would have had to be considerably more dominated by rotation, prior to mass loss, than it is now. However, we emphasise that  lack of rotation does not necessarily imply low dark matter content -- the disk may well have been dispersion dominated before infall into the cluster. 

\subsection{Can the mass loss by harassment be detected?}
A clear finding in all harassment simulations is that mass loss is substantial and unavoidable. Even the `weakly' harassed galaxy models of our sample lose over half their dark matter. This poses a question - is this mass loss detectable through observations of GCS dynamics? Perhaps by comparing a sufficient sample of field and cluster galaxies, and looking for an offset in enclosed mass?

\begin{figure}
  \centering \epsfxsize=8.5cm
  \epsffile{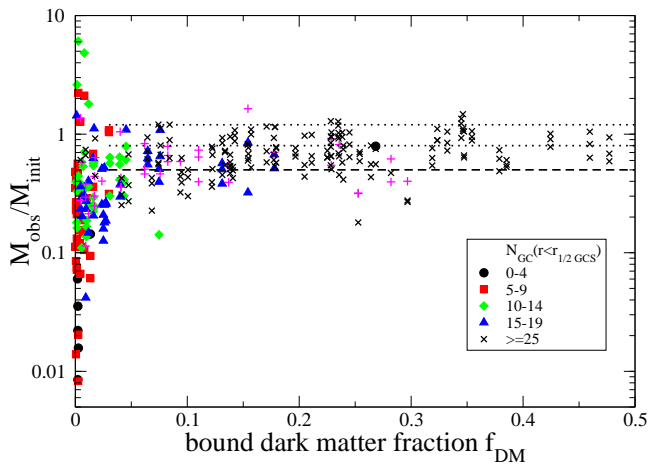}
  \caption{The ratio of the `observed' mass over the measured pre-harassment mass (measured within $r_{\frac{1}{2}{\rm{GCS}}}$),  plotted against the bound dark matter fraction. Some galaxies appear to contain equal dark matter mass as prior to harassment ($M_{\rm{obs}}/M_{\rm{init}}$=1). The dotted horizontal lines are placed at $M_{\rm{obs}}/M_{\rm{init}}$=0.8 and 1.2, while the dashed horizontal line is at $M_{\rm{obs}}/M_{\rm{init}}$=0.5. For $f_{\rm{DM}}>0.03$, where enclosed masses are well measured, the fraction of galaxies that appear unaffected by harassment (i.e. between the dotted lines) is roughly equal to galaxies that show more than half their mass has been lost (i.e. below the dashed line).}
\label{massdeficitfig}
\end{figure}

We attempt to answer this question by calculating the ratio of the `observed' mass (calculated using Equation \ref{massenclosedeqn}) over the measured, pre-harassment mass enclosed in the same $r_{\frac{1}{2}{\rm{GCS}}}$ of the GCS. We plot this against the bound dark matter fraction in Figure \ref{massdeficitfig} along three independent lines of sight, for all of our model galaxies along all 16 orbits (384 data points in total). We exclude galaxies that are completely disrupted ($f_{\rm{DM}}=f_\star$= 0.00) leaving 330 data points in total. The non-virialised state of the GCS, combined with `line-of-sight' errors, can result in a fraction of the galaxies appearing to contain equal quantities (or more) of dark matter before and after harassment. We note that it is not possible for the galaxies to truly have more dark matter than before infall - they only {\it{appear}} to have increased mass due to their GCs being supervirial, and due to unbound GCs and other projection effects. In the following we shall split our sample into galaxies that {\it{appear}} unaffected by harassment, and those which {\it{appear}} to have been significantly affected. We emphasise the word `appear' to be clear that this is observed mass we would measure using Equation \ref{massenclosedeqn}, and not necessarily their true enclosed mass (e.g. see Figure \ref{massobsfig}). The dotted lines on Figure \ref{massdeficitfig} indicate galaxies such as these, who appear almost unaffected by harassment, with an estimated enclosed mass of 0.8-1.2 of their initial mass. All galaxies below the dashed line appear heavily affected, as they appear to have lost more than half their initial mass. 

The relative number of galaxies which appear affected or appear unaffected is a function of the bound dark matter fraction that the galaxies in the sample have. We note that the GCS dynamics only probes the enclosed mass surrounding the GCs. Even when 85$\%$ of the dark matter halo has been stripped ($f_{\rm{DM}}$=0.15), typically the mass of dark matter surrounding the GCs has only fallen by roughly a factor of 2 (see Figure \ref{fcritdmfig}), where as the mass loss of stars is almost negligible ($f_\star$=0.9). Therefore measurements of the enclosed mass would not be expected to show a substantial mass deficit until $f_{\rm{DM}}<0.15$. 

If we consider only galaxies with $f_{\rm{DM}}<0.03$, galaxies that appear affected by harassment (74$\%$) outnumber galaxies that appear unaffected (4$\%$) considerably. However if we consider only galaxies with $f_{\rm{DM}}>0.03$, where the enclosed mass is well measured, galaxies that appear affected by harassment (23$\%$) are found in roughly equal numbers to galaxies that appear unaffected (26$\%$). 

In summary, in a sample of cluster dwarf elliptical galaxies, the number of galaxies that appear unaffected by harassment, compared to the number that appear to have lost half their mass, depends on the bound dark matter fractions of the galaxies in the sample. In a real galaxy sample, the bound dark matter fraction of individual galaxies might be expected to vary considerably, depending primarily on orbit but with significant scatter due to the stochastic occurrence of high speed galaxy-galaxy encounters. Galaxies that {\it{appear}} affected significantly outnumber galaxies that {\it{appear}} unaffected if there is little dark matter remaining, and galaxies are close to complete disruption. However if the galaxies have $>3\%$ of their dark matter remaining, galaxies that {\it{appear}} affected are found in roughly equal numbers to galaxies that {\it{appear}} unaffected.


\section{Summary $\&$ Conclusions}
We conduct N-body simulations of cluster early-type dwarf galaxies, surrounded by a globular cluster system (GCS). Our standard model is an approximate model for luminous Virgo early-type dwarfs, and we consider galaxy models with a range of halo mass, halo concentration, disk size, and spatial distribution of the GCS. We expect that the range of parameter space we have covered should encompass the properties of the majority of the early type dwarfs, who contain sufficient numbers of GCs to be useful for mass measurements. We evolve each galaxy model for 2.5~Gyr within a dynamic, time-evolving potential field that mimics the effects of harassment in a galaxy cluster like Virgo. Each galaxy model experiences widely ranging strengths of harassment. 

We first study the amount of dark matter, stars, and globular clusters that are stripped by harassment. We find that the stars and globular clusters are not stripped in significant quantities until the remaining bound dark matter fraction falls to 10-15$\%$, and when the dark matter density surrounding the disk has been reduced by roughly a factor of two. 

We also study the effects of harassment on the dynamics of the GCS. We find that the bound GCs are typically close to virialised (Q=0.53$\pm0.11$) within the galaxy potential while the bound dark matter fraction is above 3$\%$. We also provide a normalising constant ($\eta$), valid when there is no rotational support, that enables the enclosed mass of dwarf galaxies to be calculated from their observed GCS velocity dispersion. $\eta$ has intrinsic uncertainty of $30\%$. Including additional sources of error, that will plague pointed observations of dwarf galaxy GCSs (e.g. unbound GCs, assymetric spatial distribution of GCs, and other line-of-sight effects), we find that enclosed masses may be accurate to within a factor of 2 of their true masses. This is valid while the bound dark matter fraction is above $3\%$.

If the bound dark matter fraction is below this critical $3\%$ value, the galaxy is close to complete disruption. At this stage, bound members of the GCS can become super-virial. Unbound GCs can also be found close to the body of the galaxy resulting in a boosted velocity dispersion and therefore the GCS dynamics cannot be used as a reliable measure of the enclosed mass. It is therefore critical that we can determine if a galaxy has more or less than $3\%$ of its dark matter still bound. 

A possible indicator for whether the galaxy has $>3\%$ dark matter bound could be measurement of rotation in the stellar disk. When the bound dark matter fraction is reduced below $3\%$, typically the peak in the rotation curve is reduced to $<20\%$ of its pre-harassment value. Therefore a galaxy that currently appears to be rotationally supported is likely to contain sufficient dark matter, that dynamics of its GCS can provide an estimate of the mass enclosed, to within a factor of 2 of its real value.

Our key results may be summarised as follows.
\begin{enumerate}
\item Harassment results in the stripping of dark matter, stars and globular clusters (GCs) from our model galaxies. We consider a wide range of parameters for our model galaxies varying halo mass, and halo concentration, disk size, and globular cluster spatial distribution. We expect this parameter study to encompass the properties of most early-type dwarfs of interest for enclosed mass measurements. We find that mass loss is not very sensitive to the model galaxy parameters.
\item As dark matter is stripped, the bound dark matter fraction $f_{\rm{DM}}$ falls. We find that $f_{\rm{DM}}$ is a key parameter controlling baryonic mass loss, GC spatial distribution and GC dynamics.
\item Stars and GCs are not significantly stripped until the bound dark matter fractions is very low($f_{\rm{DM}}<$10-15$\%$), with little dependece of parameters of the model galaxies we consider.
\item Unbound GCs are often stripped to large radii. However, even considering bound GCs only, the globular cluster system (GCS) can expand spatially in response to the changing depth of the potential in which the GCs orbit. For $f_{\rm{DM}}>5\%$, the bound GCS can expand by a factor of 2-3. For $f_{\rm{DM}}<5\%$, as the galaxy approaches total disruption, the bound GCS can expand by a factor of 4-8.
\item Despite the stripping of GCs, an observed velocity dispersion, along a line-of-sight to the GCS, can be used to estimate the enclosed mass surrounding the GCS to within a factor of 2. This remains valid even when $f_{\rm{DM}}$ is as low as 3$\%$.
\item However once $f_{\rm{DM}}<3\%$, the galaxy is close to total disruption, and the observed velocity dispersion can no longer provide reliable mass estimates. Observing a rotationally dominated stellar disk could indicate that a galaxy has $f_{\rm{DM}}>3\%$.
\end{enumerate}

In this study, we have focussed on globular cluster system (GCS) dynamics as a means of measuring the enclosed mass. However in a future publication we shall study a number of other related topics including harassment-induced rotation in the GCS, the link between spatial distribution and orbit, and the fate of stripped GCs.

The tidal potential field of a galaxy cluster has a destructive influence on cluster galaxies. Substantial loss of dark matter is seen in almost all harassment simulations (\citealp{Mastropietro2005}; \citealp{Aguerri2009}; \citealp{Smith2010a}), and is almost inevitable. One reason to measure the enclosed mass of cluster dwarf elliptical galaxies might be to try to detect this mass loss. In our models, we find that the mass loss may be detectable. However we add that, within such a sample, galaxies that appear unaffected by mass loss may be equally common as galaxies that appear to have lost half their enclosed mass. Therefore a reasonable number of galaxies may be required to detect the mass-loss.

We additionally note that our findings may have implications for the assembly of larger galaxies via merging of smaller galaxies. In the cluster environment, dwarf galaxies can have $\sim85\%$ of their dark matter stripped, but keep almost all their GCs. Therefore input of dark matter through merging may be overestimated, if a `pre-harassment' globular cluster to dark matter mass ratio is assumed.

\section*{Acknowledgements}
MF acknowledges support by FONDECYT grant 1095092, RS was financed through a combination of FONDECYT grant 3120135 and a COMITE MIXTO grant, and RSJ was financed through an ESO fellowship. Funding for this research was provided in part by the Marie Curie Actions of the European Commission (FP7-COFUND).
T.H.P. acknowledges support by the FONDECYT Regular Project No. 1121005, Gemini-CONICYT Program No. 32100022, as well as support from the FONDAP Center for Astrophysics (15010003). MF and T.H.P acknowledge support from the BASAL Center for Astrophysics and Associated Technologies (PFB-06), Conicyt, Chile. JALA was supported by the projects AYA2010-21887-C04-04 and by the Consolider-Ingenio 2010 Programme grant CSD2006-00070. We thank the anonymous referee whose thorough report has significantly improved this paper.
\bibliography{bibfile}

\bsp

\label{lastpage}

\end{document}